\documentstyle[12pt]{article}

\input{epsf.sty}

\begin{document}

\begin{titlepage}
\begin{flushright}
{\large \bf UCL-IPT-96-20}
\end{flushright}
\vskip 2cm
\begin{center}

{\Large \bf
Behaviour of the Absorptive Part of the $W^{\pm }$ Electromagnetic Vertex}
\vskip 1cm

{\large M. Beuthe\footnote{IISN researcher under contract 4.4509.86}, R. Gonzalez Felipe, G. L\'{o}pez Castro\footnote{Permanent
 address: Departamento de F\'{i}sica, CINVESTAV del IPN, Apartado Postal 14-740, 
07000 M\'{e}xico D.F., M\'{e}xico}  and  J. Pestieau}\\

{\em Institut de Physique Th\'eorique, Universit\'e catholique de Louvain,}\\

{\em  B-1348 Louvain-la-Neuve, Belgium}

\end{center}

\vskip 2cm

\begin{abstract}
The absorptive part of the $WW\gamma $ vertex induced by massive fermion loops
 is considered for different kinematical configurations. We show that the axial part of 
this vertex is different from zero not only when massive fermions are involved but also
 for massless fermion loops, if one of the $W$ bosons is space-like and the other is 
time-like. We also discuss in what sense Low's soft photon theorem is satisfied.
\end{abstract}

\end{titlepage}%

\section{\protect\medskip Introduction}

\medskip

The unstable $W^{\pm }$ and $Z^{0}$ gauge bosons cannot be prepared as
asymptotic states. They are first produced in collisions of stable particles
and subsequently we observe their decay products. The intermediate states
associated to resonances in these processes are described by propagators in
the physical amplitude. The latter blows up in the resonance kinematical
region, unless an imaginary part, which softens this singularity, is added
to the propagators. The solution to the above problem in field theory
requires to include an infinite summation of vacuum polarization diagrams in
the gauge boson propagator. At a given order in perturbation theory, this
procedure can destroy the gauge invariance of the resulting amplitudes and
therefore, can induce important changes in observable effects \cite{stuart}-%
\cite{argyres}.

Different prescriptions to obtain gauge-invariant amplitudes have been
presented in the literature \cite{stuart}-\cite{pilaftsis}. In the presence
of unstable particles, this can be achieved either, (a) by making a Laurent
expansion of the full amplitude around the complex pole associated to the
resonance in the second Riemann sheet \cite
{stuart,willenbrock,wackeroth,lopez} or, (b) by using full propagators and
vertex functions in the amplitudes which include only resummation of
(gauge-invariant) fermion loops \cite{argyres}-\cite{wackeroth}. (For a
different approach, see also \cite{pilaftsis}).

In this paper we shall discuss the electromagnetic gauge invariance in
processes where a real photon is emitted from an internal $W^{\pm }$ gauge
boson. Since gauge invariance in a given process is guaranteed by the
fulfillment of the Ward identities among different Green functions, we will
focus our analysis on the latter requirement. This issue is relevant for
instance in processes involving the production and decay of a $W$ boson such
as $e^{+}e^{-}\rightarrow e^{-}\bar{\nu}_{e}ud$ \cite{argyres,lopez}, $q\bar{%
q}^{\prime }\rightarrow \ell \nu _{\ell }\gamma $ \cite{baur}, the top quark
decay $t\rightarrow b\ell \nu _{\ell }\gamma $ , or $\gamma e^{-}\rightarrow
\nu _{e}\tau ^{-}\bar{\nu}_{\tau }$ \cite{pilaftsis}.

In previous papers \cite{argyres}-\cite{wackeroth} the problem of gauge
invariance has been considered including the imaginary parts induced by
massless fermion loops in the $W^{\pm }$ propagator and the $WW\gamma $
vertex. Here we shall address this problem in the case when the fermions in
the loops are massive. Generally speaking, mass effects of light fermions in
the loops turn out to be negligible for invariant masses of the $W$ bosons
large enough. However, we would like to stress two new relevant features
that appear in both the massless and massive cases, namely, the appearance
of axial-vector contributions to the $WW\gamma $ vertex and the violation of
Low's photon energy expansion \cite{low} (see section 3) for specific
kinematical configurations of $W$'s four-momenta.

\medskip

\section{Electromagnetic gauge invariance for unstable $W^{\pm }$ bosons}

\medskip

In this section we will consider the electromagnetic Ward identity relating
the $W$ propagator and the $WW\gamma $ vertex in the one-loop approximation.
Since we are interested in gauge invariance in presence of a finite width
for the $W$ boson, we will study only the absorptive parts of the one-loop
corrections. Following Refs. \cite{argyres,baur}, we will consider the
kinematical situations where the absorptive parts of two- and three-point
functions are generated by fermion loops.

For definiteness let us consider the $W$ propagator in the unitary gauge.
Dyson resummation of vacuum polarization graphs leads to the following form
of the $W$ propagator:

\begin{equation}
D^{\mu \nu }(q)=\frac{-i}{q^{2}-M_{W}^{2}+i\mbox{Im}\,\Pi ^{T}(q^{2})}\left(
g^{\mu \nu }-\frac{q^{\mu }q^{\nu }}{q^{2}}\right) +\frac{i}{M_{W}^{2}-i%
\mbox{Im}\,\Pi ^{L}(q^{2})}\frac{q^{\mu }q^{\nu }}{q^{2}},  \label{2.1}
\end{equation}
and its inverse

\begin{equation}
\left( iD^{\mu \nu }(q)\right) ^{-1}=\left( q^{2}-M_{W}^{2}+i\mbox{Im}\,\Pi
^{T}(q^{2})\right) \left( g^{\mu \nu }-\frac{q^{\mu }q^{\nu }}{q^{2}}\right)
-\left( M_{W}^{2}-i\mbox{Im}\,\Pi ^{L}(q^{2})\right) \frac{q^{\mu }q^{\nu }}{%
q^{2}},
\end{equation}
where the transverse ($T$) and longitudinal ($L$) pieces of the $W$ vacuum
polarization are defined as:

\begin{equation}
\mbox{Im}\,\Pi ^{\alpha \beta }(q)=\mbox{Im}\,\Pi ^{T}(q^{2})\left(
g^{\alpha \beta }-\frac{q^{\alpha }q^{\beta }}{q^{2}}\right) +\mbox{Im}\,\Pi
^{L}(q^{2})\frac{q^{\alpha }q^{\beta }}{q^{2}}.  \label{2.2}
\end{equation}

\medskip

To define the $WW\gamma $ vertex we use the convention $W_{\alpha
}^{-}(q_{1})\rightarrow W_{\beta }^{-}(q_{2})+\gamma _{\mu }(k)$ , where $%
\alpha ,\beta ,\mu $ denote the four-polarization indices and $q_{1},q_{2},k$
the corresponding four-momenta. In what follows we will assume that the $W$
bosons are virtual and the photon is real. After including the one-loop
corrections (cf{\it .} Fig. 1) we obtain the following form of the full $%
WW\gamma $ vertex:

\begin{equation}
-ie\Gamma ^{\alpha \beta \mu }=-ie\left( \Gamma _{0}^{\alpha \beta \mu
}+\Gamma _{1}^{\alpha \beta \mu }\right) ,  \label{2.3}
\end{equation}
where

\begin{equation}
\Gamma _{0}^{\alpha \beta \mu }=g^{\alpha \beta }(q_{1}+q_{2})^{\mu
}-g^{\alpha \mu }(q_{1}+k)^{\beta }-g^{\beta \mu }(q_{2}-k)^{\alpha }
\label{2.4}
\end{equation}
corresponds to the tree-level expression and $\Gamma _{1}^{\alpha \beta \mu
} $ denotes the absorptive part of the one-loop correction.

The QED\ Ward identity which relates the two- and three-point functions
involving the $W$ boson is given by

\begin{equation}
k_{\mu }\Gamma ^{\alpha \beta \mu }=(iD^{\alpha \beta
}(q_{1}))^{-1}-(iD^{\alpha \beta }(q_{2}))^{-1}.  \label{2.5}
\end{equation}
This identity provides (at any order of perturbation theory) a linear
relation between the vertex correction and the vacuum polarization, which
reads for the present case

\begin{equation}
k_{\mu }\Gamma _{1}^{\alpha \beta \mu }=i\mbox{Im}\,\Pi ^{\alpha \beta
}(q_{1})-i\mbox{Im}\,\Pi ^{\alpha \beta }(q_{2}).  \label{2.6}
\end{equation}

Using the explicit expressions for these corrections we can verify the
validity of Eq.(\ref{2.6}). Indeed, for simplicity let us consider only the
contribution of the $(\nu _{\tau },\tau ^{-})$ doublet to the fermion loops.
Using the standard cutting rules, the vacuum polarization is given by

\begin{equation}
\mbox{Im}\,\Pi _{\tau }^{\alpha \beta }(q)=-\frac{g^{2}}{8}\int \frac{%
d\Omega }{32\pi ^{2}}\mbox{Tr}\,\left\{ \not{p}_{\tau }\gamma ^{\alpha }(\not%
{q}-\not{p}_{\tau })\gamma ^{\beta }(1-\gamma ^{5})\right\} \frac{%
q^{2}-m_{\tau }^{2}}{q^{2}}\theta (q^{2}-m_{\tau }^{2}),  \label{2.7}
\end{equation}
where $\theta (x)$ denotes the step function, $g$ is the $SU(2)$ weak
coupling constant, $m_{\tau }$ and $p_{\tau }^{\mu }$ are the $\tau $-lepton
mass and four-momentum respectively.

On the other hand, the vertex correction contains two terms corresponding to
two different cut diagrams depicted in Figs. (1a) and (1b), namely,

\begin{equation}
\Gamma _{1}^{\alpha \beta \mu }=i\left( I_{\tau }^{\alpha \beta \mu }(q_{1})+%
\widetilde{I}_{\tau }^{\alpha \beta \mu }(q_{2})\right) ,  \label{2.8}
\end{equation}
where

\begin{eqnarray}
I_{\tau }^{\alpha \beta \mu }(q_{1}) &=&\frac{Q_{\tau }g^{2}}{8}\int \frac{%
d\Omega }{32\pi ^{2}}\, \frac{q_{1}^{2}-m_{\tau }^{2}}{2(k\cdot p_{\tau
})q_{1}^{2}}\, \theta (q_{1}^{2}-m_{\tau }^{2})  \label{2.9} \\
&&\times \mbox{Tr}\,\left\{ \not{p}_{\tau }\gamma ^{\alpha }(\not{q}_{1}-\not%
{p}_{\tau })\gamma ^{\beta }(2p_{\tau }^{\mu }-\not{k}\gamma ^{\mu
})(1-\gamma ^{5})\right\} ,  \nonumber
\end{eqnarray}

\begin{eqnarray}
\widetilde{I}_{\tau }^{\alpha \beta \mu }(q_{2}) &=&-\frac{Q_{\tau }g^{2}}{8}%
\int \frac{d\Omega }{32\pi ^{2}}\, \frac{q_{2}^{2}-m_{\tau }^{2}}{2(k\cdot
p_{\tau })q_{2}^{2}}\, \theta (q_{2}^{2}-m_{\tau }^{2})  \label{2.10} \\
&&\times \mbox{Tr}\,\left\{ \not{p}_{\tau }(2p_{\tau }^{\mu }+\gamma ^{\mu }%
\not{k})\gamma ^{\alpha }(\not{q}_{2}-\not{p}_{\tau })\gamma ^{\beta
}(1-\gamma ^{5})\right\} ;  \nonumber
\end{eqnarray}
$Q_{\tau }=-1$ is the $\tau $ - lepton electric charge in units of the
positron charge $e$.

It is straightforward to check that the QED Ward identity is satisfied when
we insert Eqs.(\ref{2.7})-(\ref{2.10}) into Eq.(\ref{2.6}). In other words,
the electromagnetic gauge invariance is guaranteed in a process involving
the production and decay of $W$ bosons, if Eqs.(\ref{2.1}) and (\ref{2.3})
are used for the $W$ propagator and $WW\gamma $ vertex function,
respectively.

Let us now consider the relationship between the above results and those of
Ref.\cite{lopez}. If the fermions in the loops are massless then $\mbox{Im}%
\,\Pi ^{L}(q^{2})=0$, Im$\,\Pi ^{T}(q^{2})=q^{2}\Gamma _{W}/M_{W}=\sum
g^{2}q^{2}/48\pi $ (the sum runs over doublets and colours) and we obtain 
\cite{baur} when $q^{2}>0:$

\begin{equation}
D^{\mu \nu }(q)=\frac{-i}{q^{2}-M_{W}^{2}+iq^{2}\gamma _{W}}\left( g^{\mu
\nu }-\frac{q^{\mu }q^{\nu }}{M_{W}^{2}}(1+i\gamma _{W})\right) ,
\label{2.11}
\end{equation}
where $\gamma _{W}\equiv \Gamma _{W}/M_{W}$ and $\Gamma _{W}$ denotes the
total width of the $W$ boson for massless fermions. In this case the $%
WW\gamma $ vertex receives only vector contributions and we have \cite{baur}
(see also Eqs.(\ref{3.1}),(\ref{3.3}) and (\ref{3.10}) below)

\begin{equation}
-ie\Gamma ^{\alpha \beta \mu }=-ie\left( 1+i\gamma _{W}\right) \Gamma
_{0}^{\alpha \beta \mu }.  \label{2.12}
\end{equation}

Note that the propagator (\ref{2.11}) can be rewritten as follows:

\begin{equation}
D^{\mu \nu }(q)=\frac{-i(1+i\overline{\gamma }_{W})^{-1}}{q^{2}-\overline{M}%
_{W}^{2}+i\overline{M}_{W}\overline{\Gamma }_{W}}\left( g^{\mu \nu }-\frac{%
q^{\mu }q^{\nu }}{\overline{M}_{W}^{2}-i\overline{M}_{W}\overline{\Gamma }%
_{W}}\right) ,  \label{2.13}
\end{equation}
with $\overline{M}_{W}^{2}=M_{W}^{2}+\Gamma _{W}^{2}$ and $\overline{\gamma }%
_{W}\equiv \overline{\Gamma }_{W}/\overline{M}_{W}=\Gamma _{W}/M_{W}.$

The form of the propagator given in Eq.(\ref{2.13}) was obtained in \cite
{lopez} using a Laurent expansion of the full propagator around the complex
pole position $\overline{M}_{W}^{2}-i\overline{M}_{W}\overline{\Gamma }_{W}$%
, where $(1+i\overline{\gamma }_{W})^{-1}$ plays the role of a (complex)
renormalization wave function for the $W$ boson. Thus, the form of the
propagator derived in Ref.\cite{lopez}, which is valid near the resonance,
is identical to the full $W$ propagator in the limit of massless fermions in
loop corrections. Since the full $WW\gamma $ vertex in the massless case has
the same tensor structure as the tree-level expression\footnote{%
Indeed, if we keep the non-leading renormalization wave function factor in
Eq.(14), the Ward identity forces the $WW\gamma $ vertex to have the form
given in Eq.(13).}\medskip , the Ward identity is also satisfied. We note
that $\left| 1+i\gamma _{W}\right| ^{2}\simeq 1.0007$ is a modest correction.

\section{\protect\medskip Absorptive corrections to $WW\gamma $ vertex:
massless\protect\linebreak fermions}

\medskip

In this section we consider explicitly the absorptive part of the one-loop
corrections to the $WW\gamma $ vertex (Fig. 1) in the case of massless
fermions (except for the top quark). We will focus our attention on the
structure of these vertex corrections for two different kinematical
configurations involved in processes of current interest, namely, (a) $%
q_{1}^{2}<0,q_{2}^{2}>0$ and (b) $q_{1}^{2}>0,q_{2}^{2}>0.$ In particular,
we will discuss the behaviour of the corrections {\em in the soft photon
limit }and the non-zero contribution to the axial-vector part induced by the
decoupling of the top quark from the absorptive vertex corrections.

The momenta in the $WW\gamma $ vertex flow according to $W_{\alpha
}^{-}(q_{1})\rightarrow W_{\beta }^{-}(q_{2})+\gamma _{\mu }(k),$ with $%
q_{1}=q_{2}+k$ and the photon being on-shell. Since the top quark ($%
m_{t}=175\pm 6$ GeV) \cite{blondel} is far from being massless, the doublet
of fermions $(t,b)$ containing the top quark does not contribute to
absorptive corrections when $q_{1,2}^{2}<m_{t}^{2}$. In this case, the
vertex corrections can be written as

\begin{equation}
\Gamma _{1}^{\alpha \beta \mu }=ic\left\{ \left( 3Q_{L}+6Q_{D}-6Q_{U}\right) 
\overline{V}^{\alpha \beta \mu }+i\left( 3Q_{L}+6Q_{D}+6Q_{U}\right) 
\overline{A}^{\alpha \beta \mu }\right\} ,  \label{3.1}
\end{equation}
where $Q_{i}$ $(i=L,U,D)$ denote the electric charges of leptons and quarks
in units of the positron charge $e$ and
\begin{equation}
c\equiv -\frac{g^{2}}{64\pi (k\cdot q_{1})}.  \label{3.2}
\end{equation}

The vector and axial-vector contributions in Eq.(\ref{3.1}) are given,
respectively, by

\begin{equation}
\overline{V}^{\alpha \beta \mu }=V_{f}^{\alpha \beta \mu }(q_{1})\,\theta
(q_{1}^{2})+\widetilde{V}_{f}^{\alpha \beta \mu }(q_{2})\,\theta (q_{2}^{2}),
\label{3.3}
\end{equation}

\begin{equation}
\overline{A}^{\alpha \beta \mu }=A_{f}^{\alpha \beta \mu }(q_{1})\,\theta
(q_{1}^{2})+\widetilde{A}_{f}^{\alpha \beta \mu }(q_{2})\,\theta (q_{2}^{2}),
\label{3.4}
\end{equation}
where $f$ refers to the fermion that emits the photon.

Note that if the top contributed to the absorptive corrections $%
(q_{1}^{2},q_{2}^{2}\gg m_{t}^{2})$, we would have to add the term $%
3ic(Q_{D}-Q_{U})V^{\alpha \beta \mu }-3c(Q_{D}+Q_{U})A^{\alpha \beta \mu }$
to the r.h.s. of Eq.(\ref{3.1}). The coefficient of the axial-vector term
would become $-3c(Q_{L}+3Q_{D}+3Q_{U})=0,$ which assures the vanishing of
this contribution as noticed in \cite{argyres}.

Using the results given in the Appendix we can write the explicit
expressions for the terms in Eqs.(\ref{3.3}),(\ref{3.4}) in the limit where
all fermions except the $t$-quark have zero mass:

\begin{eqnarray}
V_{f}^{\alpha \beta \mu }(q_{1}) &=&\frac{4}{3}\left( q_{1}^{2}g^{\alpha
\beta }-q_{1}^{\alpha }q_{1}^{\beta }\right) q_{1}^{\mu }-2q_{1}^{2}\left(
k^{\beta }g^{\alpha \mu }-k^{\alpha }g^{\beta \mu }\right)  \nonumber \\
&&+\frac{1}{3}\left( q_{1}^{2}q_{1}^{\alpha }-\frac{q_{1}^{4}}{k\cdot q_{1}}%
k^{\alpha }-6q_{1}^{\alpha }(k\cdot q_{1})\right) \left( g^{\beta \mu }-%
\frac{k^{\beta }q_{1}^{\mu }}{k\cdot q_{1}}\right)  \nonumber \\
&&+\frac{1}{3}\left( q_{1}^{2}q_{1}^{\beta }-\frac{q_{1}^{4}}{k\cdot q_{1}}%
k^{\beta }+3k^{\beta }q_{1}^{2}\right) \left( g^{\alpha \mu }-\frac{%
k^{\alpha }q_{1}^{\mu }}{k\cdot q_{1}}\right) ,  \label{3.5}
\end{eqnarray}

\begin{equation}
\widetilde{V}_{f}^{\alpha \beta \mu }(q_{2})=V_{f}^{\beta \alpha \mu
}(-q_{2}),  \label{3.6}
\end{equation}

\begin{equation}
A_{f}^{\alpha \beta \mu }(q_{1})=q_{1}^{2}\varepsilon ^{\alpha \beta \mu
\sigma }(q_{1}-k)_{\sigma }+2\varepsilon ^{\beta \mu \rho \sigma }q_{1\rho
}k_{\sigma }q_{1}^{\alpha }+\frac{q_{1}^{2}}{k\cdot q_{1}}\left( \varepsilon
^{\alpha \beta \rho \sigma }q_{1}^{\mu }-\varepsilon ^{\beta \mu \rho \sigma
}k^{\alpha }\right) q_{1\rho }k_{\sigma },  \label{3.7}
\end{equation}

\begin{equation}
\widetilde{A}_{f}^{\alpha \beta \mu }(q_{2})=-A_{f}^{\beta \alpha \mu
}(-q_{2}).  \label{3.8}
\end{equation}

The above corrections satisfy the following identities:

\begin{eqnarray}
k_{\mu }V_{f}^{\alpha \beta \mu }(q_{1}) &=&\frac{4}{3}(k\cdot q_{1})\left(
q_{1}^{2}g^{\alpha \beta }-q_{1}^{\alpha }q_{1}^{\beta }\right) ,  \nonumber
\\
k_{\mu }\widetilde{V}_{f}^{\alpha \beta \mu }(q_{2}) &=&-\frac{4}{3}(k\cdot
q_{2})\left( q_{2}^{2}g^{\alpha \beta }-q_{2}^{\alpha }q_{2}^{\beta }\right)
,  \label{3.9} \\
k_{\mu }A_{f}^{\alpha \beta \mu }(q_{1}) &=&k_{\mu }\widetilde{A}%
_{f}^{\alpha \beta \mu }(q_{2})=0.  \nonumber
\end{eqnarray}

\medskip Let us first consider the case $q_{1}^{2}>0,q_{2}^{2}>0.$ This
kinematical situation is present, for instance, in the process $q\bar{q}%
^{\prime }\rightarrow \ell \nu _{\ell }\gamma $ \cite{baur,wackeroth}. In
this case both cuts in Eqs.(\ref{3.3}),(\ref{3.4}) contribute and we get
\begin{eqnarray}
\overline{V}^{\alpha \beta \mu } &=&\frac{4}{3}(k\cdot q_{1})\Gamma
_{0}^{\alpha \beta \mu },  \nonumber \\
\overline{A}^{\alpha \beta \mu } &=&0,  \label{3.10}
\end{eqnarray}
where $\Gamma _{0}^{\alpha \beta \mu }$ is the tree-level $WW\gamma $ vertex
defined in Eq.(\ref{2.4}).

If we insert the above expressions into Eq.(\ref{3.1}) we recover Eq.(\ref
{2.12}) {\it i.e.} the result of \cite{baur}, namely, one-loop corrections
to the $WW\gamma $ vertex accounts for the rescaling of the tree-level
expression by a factor $1+i\Gamma _{W}/M_{W}.$ Notice however that this
factorization of the $WW\gamma $ vertex is valid only when $q_{1,2}^{2}>0.$

Observe that the axial contribution vanishes when we add the two cuts
associated to the photon emission from a given fermionic line (cf. Eqs. (\ref
{3.4}),(\ref{3.7}),(\ref{3.8})). In other words, when $%
q_{1}^{2}>0,q_{2}^{2}>0,$ the axial-vector contribution vanishes, regardless
of the condition for the cancellation of anomalous terms, $%
Q_{L}+3Q_{D}+3Q_{U}=0$.

A second comment concerns the behaviour of vertex corrections in the soft
photon limit. According to the Low's soft-photon theorem \cite{low}, the
leading terms (of order $\omega ^{-1}$ and $\omega ^{0}$ , $\omega $ is the
photon energy) in the amplitude for the emission of a soft photon in a
radiative process are determined by the electromagnetic gauge invariance and
the corresponding non-radiative process. Terms of the form $O(\omega ^{-1})$
arise only from photon emission off external charged particles, while photon
emission off internal lines starts at order $O(\omega ^{0})$ and cannot be
of the form $k^{\alpha }/k\cdot q$ ($k$ is the photon momentum and $q$ is
the momentum of an internal line). Now, if we insert Eq.(\ref{3.10}) into
Eq.(\ref{3.1}) we can easily check that terms proportional to $k\cdot q_{1}$
drop and the vector contribution satisfies in this case Low's theorem. As we
shall see in the next section, this property remains valid in the massive
case for kinematical configurations when all possible cuts (for the photon
emission from a given fermionic line) are allowed.

Let us now consider the case $q_{1}^{2}<0,q_{2}^{2}>0.$ Such kinematical
configuration is present {\it e.g. }in the process $\gamma e^{-}\rightarrow
\nu _{e}\tau ^{-}\bar{\nu}_{\tau }$ \cite{pilaftsis} with the appropriate
change of the photon momentum $k\rightarrow -k.$ According to Eqs.(\ref{3.3}%
),(\ref{3.4}) only one cut diagram for each loop (Figs. 1b, 1d, 1f) is
allowed in this case and we obtain from Eq.(\ref{3.1}):

\begin{equation}
\Gamma _{1}^{\alpha \beta \mu }=ic\left\{ \left( 3Q_{L}+6Q_{D}-6Q_{U}\right) 
\widetilde{V}_{f}^{\alpha \beta \mu }(q_{2})+iQ_{\tau }\widetilde{A}%
_{f}^{\alpha \beta \mu }(q_{2})\right\} ,  \label{3.11}
\end{equation}
where we have used the fact that $Q_{L}+3Q_{U}+3Q_{D}=0$. Therefore, the
axial contribution is not zero in contradiction with what is claimed in Ref.%
\cite{argyres}. This term does not vanish because the heavy top quark
decouples from the absorptive vertex corrections and leaves a net
contribution coming from leptons of the third generation.

A second relevant feature in the case $q_{1}^{2}<0,q_{2}^{2}>0$ concerns the
validity of Low's expansion with respect to $k$. As can be easily realized
by inserting Eqs.(\ref{3.6}),(\ref{3.8}) into Eq.(\ref{3.11}), the vector
and axial contributions that appear in the latter equation contain terms of
order $(k\cdot q_{2})^{-1}$ and $k^{\alpha }/(k\cdot q_{2})$. This is at
variance with Low's expansion with respect to $k.$ Of course, Low's soft
photon theorem \cite{low} is not violated because it is impossible to have $%
k\rightarrow 0$ when $q_{1}^{2}<0$ and $q_{2}^{2}>0.$

Finally we note that, when $q_{1}^{2}<0,$ the factorization appearing in Eq.(%
\ref{2.12}) does not remain valid.

\section{Absorptive corrections to $WW\gamma $ vertex: massive\protect%
\linebreak fermions}

\medskip

In this section we derive the expressions for the vector and axial-vector
contributions to the absorptive part of the $WW\gamma $ vertex when the
masses of the fermions are taken into account in the loops. To illustrate
our work, we will assume that $m_{t}^{2}>q_{1,2}^{2}>m_{\tau }^{2}$ and that
only the $t,b$ and $c$  quarks and the $\tau $ lepton are massive fermions.
All the possible cut diagrams giving contributions to the absorptive part of
the $WW\gamma $ vertex are shown in Fig. 1 and the general expressions for
such contributions are provided in the Appendix for the case of arbitrary
fermion masses in the loops.

The correction to the vertex can be split into vector and axial-vector terms,

\begin{equation}
\Gamma _{1}^{\alpha \beta \mu }=ic\left( V^{\alpha \beta \mu }+iA^{\alpha
\beta \mu }\right) ,  \label{4.1}
\end{equation}
with

\begin{eqnarray}
V^{\alpha \beta \mu } &=&\sum_{L=e,\mu ,\tau }Q_{L}\left( V_{L}^{\alpha
\beta \mu }(q_{1})+\widetilde{V}_{L}^{\alpha \beta \mu }(q_{2})\right)
+3\sum_{D=s,d}Q_{D}\left( V_{D}^{\alpha \beta \mu }(q_{1})+\widetilde{V}%
_{D}^{\alpha \beta \mu }(q_{2})\right)  \nonumber \\
&&-3\sum_{U=u,c}Q_{U}\left( V_{U}^{\alpha \beta \mu }(q_{1})+\widetilde{V}%
_{U}^{\alpha \beta \mu }(q_{2})\right)  \label{4.2}
\end{eqnarray}
and

\begin{eqnarray}
A^{\alpha \beta \mu } &=&\sum_{L=e,\mu ,\tau }Q_{L}\left( A_{L}^{\alpha
\beta \mu }(q_{1})+\widetilde{A}_{L}^{\alpha \beta \mu }(q_{2})\right)
+3\sum_{D=s,d}Q_{D}\left( A_{D}^{\alpha \beta \mu }(q_{1})+\widetilde{A}%
_{D}^{\alpha \beta \mu }(q_{2})\right)  \nonumber \\
&&+3\sum_{U=u,c}Q_{U}\left( A_{U}^{\alpha \beta \mu }(q_{1})+\widetilde{A}%
_{U}^{\alpha \beta \mu }(q_{2})\right) ,  \label{4.3}
\end{eqnarray}
For simplicity we have neglected the Kobayashi-Maskawa mixing. Note also
that the two different cuts for the emission of the photon from a given
fermionic line are determined by $q_{1}^{2}$ and $q_{2}^{2}$, such that $%
q_{1}^{2}-q_{2}^{2}=2(k\cdot q_{1})=2(k\cdot q_{2})$ .

Let us first consider the vector contributions. The expressions for the
coefficients $V_{f}^{\alpha \beta \mu }(q_{1})$ and $\widetilde{V}%
_{f}^{\alpha \beta \mu }(q_{2})$ $(f=L,U,D)$ are given in the Appendix for
the general case when all the fermion masses are different from zero. Here
we will explicitly assume $m_{e}=m_{\mu }=m_{u}=m_{d}=m_{s}=0,$ but $m_{c}$
and $m_{\tau }$ are kept finite. Using the Eqs.(\ref{A1}) and (\ref{A2})
given in the Appendix we obtain

\begin{equation}
V^{\alpha \beta \mu }=\left( 3Q_{L}+6Q_{D}-6Q_{U}\right) \frac{4}{3}(k\cdot
q_{1})\Gamma _{0}^{\alpha \beta \mu }+Q_{\tau }\widehat{V}_{\tau }^{\alpha
\beta \mu }+3Q_{s}\widehat{V}_{s}^{\alpha \beta \mu }-3Q_{c}\widehat{V}%
_{c}^{\alpha \beta \mu }  \label{4.4}
\end{equation}
where $\widehat{V}_{\tau }^{\alpha \beta \mu },\widehat{V}_{s}^{\alpha \beta
\mu }$ and $\widehat{V}_{c}^{\alpha \beta \mu }$ correspond to the photon
emission from the $\tau $ lepton and the $s$ and $c$ quarks, respectively,
and they vanish in the limit of massless fermions. The explicit expressions
for these terms are

\begin{eqnarray}
\widehat{V}_{\tau }^{\alpha \beta \mu } &=&\left\{ -\frac{2m_{\tau
}^{6}(k\cdot q_{1})(q_{1}^{2}+q_{2}^{2})}{3q_{1}^{4}q_{2}^{4}}g^{\alpha
\beta }q_{1}^{\mu }-2\left[ -\frac{m_{\tau }^{2}}{\ k\cdot q_{1}}\ln \left( 
\frac{q_{1}^{2}}{q_{2}^{2}}\right) +\frac{2m_{\tau }^{4}(k\cdot
q_{1})(2q_{2}^{2}-q_{1}^{2})}{q_{1}^{4}q_{2}^{4}}\right. \right.  \nonumber
\\
&&\left. -\frac{4m_{\tau }^{6}(k\cdot q_{1})(q_{1}^{2}+2q_{2}^{2})}{%
3q_{1}^{6}q_{2}^{4}}+\frac{2m_{\tau }^{2}}{q_{2}^{2}}\right] q_{1}^{\alpha
}q_{1}^{\beta }q_{1}^{\mu }-\left[ 4m_{\tau }^{2}+\frac{3m_{\tau }^{4}}{%
q_{1}^{2}}+\frac{m_{\tau }^{4}}{q_{2}^{2}}+\frac{4m_{\tau }^{6}(k\cdot q_{1})%
}{3q_{1}^{4}q_{2}^{2}}\right.  \nonumber \\
&&\left. -\frac{2m_{\tau }^{2}(q_{2}^{2}+m_{\tau }^{2})}{\ k\cdot q_{1}}\ln
\left( \frac{q_{1}^{2}}{q_{2}^{2}}\right) +\frac{2m_{\tau }^{2}(k\cdot
q_{1})(m_{\tau }^{2}-2q_{1}^{2})}{\ q_{1}^{4}}\right] k^{\alpha }\left(
g^{\beta \mu }-\frac{k^{\beta }q_{1}^{\mu }}{k\cdot q_{1}}\right)  \nonumber
\\
&&+2(k\cdot q_{1})\left[ \frac{2m_{\tau }^{2}}{q_{2}^{2}}-\frac{m_{\tau }^{2}%
}{\ k\cdot q_{1}}\ln \left( \frac{q_{1}^{2}}{q_{2}^{2}}\right) +\frac{%
m_{\tau }^{4}(3q_{2}^{2}-q_{1}^{2})}{q_{1}^{2}q_{2}^{4}}-\frac{2m_{\tau
}^{6}(q_{1}^{2}+q_{2}^{2})}{3q_{1}^{4}q_{2}^{4}}\right] g^{\alpha \mu
}q_{1}^{\beta }  \nonumber \\
&&\left. \left. \left. +\frac{2m_{\tau }^{2}(k\cdot q_{1})}{%
q_{1}^{4}q_{2}^{4}}\left[ 2q_{1}^{2}q_{2}^{2}-m_{\tau
}^{2}(q_{1}^{2}+q_{2}^{2})\right] q_{2}^{\alpha }q_{1}^{\beta }q_{1}^{\mu
}\right\} +\right\{ \alpha \leftrightarrow \beta ,q_{1}\leftrightarrow
-q_{2}\right\}  \label{4.5}
\end{eqnarray}
and

\begin{eqnarray}
\widehat{V}_{s}^{\alpha \beta \mu } &=&\frac{4m_{c}^{4}(k\cdot q_{1})}{%
3q_{1}^{4}q_{2}^{4}}\left\{ -m_{c}^{2}(q_{1}^{2}+q_{2}^{2})g^{\alpha \beta
}q_{1}^{\mu }-\frac{(q_{1}^{2}+2q_{2}^{2})(3q_{1}^{2}-2m_{c}^{2})}{q_{1}^{2}}%
q_{1}^{\alpha }q_{1}^{\beta }q_{1}^{\mu }\right.  \nonumber \\
&&-\frac{(q_{2}^{2}+2q_{1}^{2})(3q_{2}^{2}-2m_{c}^{2})}{q_{2}^{2}}%
q_{2}^{\alpha }q_{2}^{\beta }q_{2}^{\mu }+\frac{3}{2}\left(
q_{2}^{4}q_{2}^{\alpha }g^{\beta \mu }+q_{1}^{4}q_{1}^{\beta }g^{\alpha \mu
}\right)  \nonumber \\
&&+\frac{1}{2}\left(
3q_{1}^{2}q_{2}^{2}-2m_{c}^{2}(q_{1}^{2}+q_{2}^{2})\right) \left(
q_{2}^{\alpha }g^{\beta \mu }+q_{1}^{\beta }g^{\alpha \mu }\ \right)
+3(q_{1}^{2}+q_{2}^{2})q_{2}^{\alpha }q_{1}^{\beta }q_{1}^{\mu }  \nonumber
\\
&&\left. +\frac{1}{2}\left( 2m_{c}^{2}-3q_{1}^{2}-3q_{2}^{2}\right) \left(
q_{1}^{2}k^{\beta }g^{\alpha \mu }-q_{2}^{2}k^{\alpha }g^{\beta \mu
}-2k^{\alpha }k^{\beta }q_{1}^{\mu }\right) \right\}  \label{4.6}
\end{eqnarray}
The expression for $\widehat{V}_{c}^{\alpha \beta \mu }$ is obtained from $%
\widehat{V}_{\tau }^{\alpha \beta \mu }$ by replacing $m_{\tau }\rightarrow
m_{c}.$

Next we discuss the vertex corrections arising from axial-vector
contributions. In our case of interest, where $m_{e}=m_{\mu
}=m_{u}=m_{d}=m_{s}=0$ but $m_{c}$ , $m_{\tau }$ different from zero, the
explicit expression for the total axial-vector contribution becomes (cf.
Appendix)

\begin{eqnarray}
A^{\alpha \beta \mu } &=&2\left\{ \varepsilon ^{\alpha \beta \mu \sigma
}k_{\sigma }\left( Q_{\tau }B_{\tau }+3Q_{c}B_{c}\right) -\varepsilon
^{\beta \mu \rho \sigma }q_{1\rho }k_{\sigma }q_{1}^{\alpha }\left( Q_{\tau
}A_{1\tau }+3Q_{s}D_{1c}+3Q_{c}A_{1c}\right) \right.  \nonumber \\
&&\left. +\varepsilon ^{\alpha \mu \rho \sigma }q_{2\rho }k_{\sigma
}q_{2}^{\beta }\left( Q_{\tau }A_{2\tau }+3Q_{s}D_{2c}+3Q_{c}A_{2c}\right)
\right\} ,  \label{4.7}
\end{eqnarray}
where ($f$ denotes $\tau $ or $c$) 
\begin{eqnarray}
A_{1f} &=&-\frac{m_{f}^{2}}{k\cdot q_{1}}\ln \left( \frac{q_{1}^{2}}{%
q_{2}^{2}}\right) +\frac{2m_{f}^{2}}{q_{1}^{2}}+D_{1f} ,  \nonumber \\
A_{2f} &=&A_{1f}\;(q_{1}\leftrightarrow -q_{2}) ,  \nonumber \\
D_{1f} &=&-\frac{2m_{f}^{4}(k\cdot q_{1})}{q_{1}^{4}q_{2}^{2}} ,  \label{4.8}
\\
D_{2f} &=&D_{1f}\;(q_{1}\leftrightarrow -q_{2}) ,  \nonumber \\
B_{f} &=&q_{1}^{2}A_{1f}+q_{2}^{2}A_{2f}\;.  \nonumber
\end{eqnarray}

Finally, let us comment that our expressions for the vector contributions to
the $WW\gamma $ vertex corrections (cf. Eqs.(\ref{4.4})-(\ref{4.6})) satisfy
the QED Ward identity given in Eq.(\ref{2.6}). Note also that the
axial-vector piece of the vertex corrections (cf. Eqs.(\ref{4.7})-(\ref{4.8}%
)) is gauge-invariant by itself and therefore does not contribute to this
Ward identity. Our results are in agreement with Low's soft photon theorem 
\cite{low}. Indeed, the expansion of Eqs.(\ref{4.5}),(\ref{4.6}) and (\ref
{4.8}) around $(k\cdot q_{1})=0$ yields an overall factor $(k\cdot q_{1})$,
which cancels the $(k\cdot q_{1})^{-1}$ factor present in the coefficient $c$
(Eq.(\ref{3.2})) of Eq.(\ref{4.1}).

\medskip

\section{Conclusions}

\medskip

The amplitude for a radiative process involving production and decay of an
unstable $W^{\pm }$ boson ({\it e.g. }$q\bar{q}^{\prime }\rightarrow \ell
\nu _{\ell }\gamma ,e^{+}e^{-}\rightarrow e^{-}\bar{\nu}_{e}ud,$ $%
t\rightarrow b\ell \nu _{\ell }\gamma $ or $\gamma e^{-}\rightarrow \nu
_{e}\tau ^{-}\bar{\nu}_{\tau })$ can be made gauge-invariant by resuming the
vacuum polarization of the $W$ and of the vertex $WW\gamma $ corrections
induced by loops of fermions \cite{argyres}-\cite{pilaftsis}. In this paper
we have explicitly shown that the absorptive corrections indeed satisfy the
corresponding electromagnetic Ward identity in the general case when loops
with massive fermions are considered.

We have also computed the absorptive parts of the $WW\gamma $ vertex
generated by massive fermions in the one-loop corrections. Contrary to the
case of massless fermions, where vertex corrections account for a simple
rescaling of the tree-level $WW\gamma $ vertex \cite{argyres,baur} when $%
q_{1}^{2}>0,q_{2}^{2}>0$ (see Eq.(\ref{2.12})), we find that the Lorentz
tensor structure of this vertex is sensibly different. In particular,
additional corrections to the magnetic dipole and electric quadrupole
moments of $W$ are induced, as clearly seen from Eqs.(\ref{4.4})-(\ref{4.6}).

The axial-vector structure in the $WW\gamma $ vertex is a novel feature in
the case of loops with massive fermions as well as in the case of massless
fermions for specific kinematical configurations of the $W$'s four-momenta
in the vertex. Such a structure induces non-zero electric dipole and
magnetic quadrupole moments of the $W$ gauge boson (see Eqs.(\ref{4.7})-(\ref
{4.8})).

Our results for the vector and axial-vector pieces of the $WW\gamma $ vertex
have the correct threshold behaviour (they vanish when $q_{1,2}^{2}%
\rightarrow (m_{f}+m_{f^{\prime }})^{2}$ ) and are in agreement with Low's
soft photon theorem \cite{low}, when the two cuts are allowed, {\it i.e.} $%
q_{1,2}^{2}>(m_{f}+m_{f^{\prime }})^{2}$ , for the emission of the photon
from a given fermionic line. When $q_{1}^{2}<0$ and $q_{2}^{2}>0,$ Low's
photon energy expansion is not satisfied and the axial-vector structure in
the $WW\gamma $ vertex is not zero even in the massless fermion case.

\bigskip 

\noindent {\bf Acknowledgements}

\medskip 

We thank Jean-Marc G\'{e}rard and Jacques Weyers for valuable discussions.

\newpage

\noindent {\sc {\bf Appendix}}

\setcounter{equation}{0} \renewcommand{\theequation}{A\arabic{equation}}

\medskip

In this Appendix we provide the general expressions for the vector and
axial-vector contributions to the absorptive part of the $WW\gamma $ vertex
in the case of massive fermions. As is well known, we have to include all
cut diagrams allowed by kinematics with on-shell fermions in the loops.

\medskip

\noindent {\rm (i)} {\em Vector contributions}

\medskip

If we consider the cut diagrams with $q_{1}$ (Figs. 1a, 1c and 1e) we obtain
the following expressions for the coefficients $V_{f}^{\alpha \beta \mu
}(q_{1})$ in Eq.(\ref{4.2}) ($f=L,U,D)$ :

\begin{eqnarray}
V_{f}^{\alpha \beta \mu }(q_{1}) &=&\frac{8E_{1}\upsilon _{1}}{\sqrt{%
q_{1}^{2}}}\left[ -\frac{2E_{1}^{2}\upsilon _{1}^{2}}{3}+E_{1}\sqrt{q_{1}^{2}%
}-m_{f}^{2}\right] g^{\alpha \beta }q_{1}^{\mu }  \nonumber \\
&&+\frac{16E_{1}^{2}\upsilon _{1}}{q_{1}^{2}\sqrt{q_{1}^{2}}}\left[
E_{1}\left( 1+\frac{\upsilon _{1}^{2}}{3}\right) -\sqrt{q_{1}^{2}}\right]
q_{1}^{\alpha }q_{1}^{\beta }q_{1}^{\mu }+\frac{8E_{1}\upsilon _{1}}{%
q_{1}^{2}}\left( E_{1}-\sqrt{q_{1}^{2}}\right) q_{1}^{\alpha }T^{\beta \mu }
\nonumber \\
&&+\frac{2}{k\cdot q_{1}}\left( -\frac{8E_{1}^{3}\upsilon _{1}^{3}}{3\sqrt{%
q_{1}^{2}}}-m_{f}^{2}\ln \left( \frac{1+\upsilon _{1}}{1-\upsilon _{1}}%
\right) +2\upsilon _{1}E_{1}^{2}\right) \left( q_{1}^{\alpha }T^{\beta \mu
}+q_{1}^{\beta }T^{\alpha \mu }\right)  \nonumber \\
&&-\frac{4E_{1}\sqrt{q_{1}^{2}}}{(k\cdot q_{1})^{2}}\left( -\frac{%
4E_{1}^{2}\upsilon _{1}^{3}}{3}-m_{f}^{2}\ln \left( \frac{1+\upsilon _{1}}{%
1-\upsilon _{1}}\right) +2\upsilon _{1}E_{1}^{2}\right) \left( k^{\alpha
}T^{\beta \mu }+k^{\beta }T^{\alpha \mu }\right)  \nonumber \\
&&-\frac{2}{k\cdot q_{1}}\left( m_{f}^{2}\ln \left( \frac{1+\upsilon _{1}}{%
1-\upsilon _{1}}\right) -2\upsilon _{1}E_{1}\sqrt{q_{1}^{2}}\right)
k^{\alpha }T^{\beta \mu }+\frac{4E_{1}\upsilon _{1}}{k\cdot q_{1}}\left(
E_{1}-\sqrt{q_{1}^{2}}\right) k^{\beta }T^{\alpha \mu } .  \nonumber \\
&&  \label{A1}
\end{eqnarray}

Similarly, the cut diagrams characterized by $q_{2}$ (Figs. 1b, 1d, 1f) give
rise to the following expression for the coefficients $\widetilde{V}%
_{f}^{\alpha \beta \mu }(q_{2})$ in Eq.(\ref{4.2}):

\begin{equation}
\widetilde{V}_{f}^{\alpha \beta \mu }(q_{2})=V_{f}^{\beta \alpha \mu
}(-q_{2}).  \label{A2}
\end{equation}
In the above equations the following definitions have been introduced:

\begin{eqnarray}
E_{1,2} &=&\frac{q_{1,2}^{2}+m_{f}^{2}-m_{f^{\prime }}^{2}}{2\sqrt{%
q_{1,2}^{2}}},\qquad \upsilon _{1,2}=\sqrt{1-\frac{m_{f}^{2}}{E_{1,2}^{2}}},
\nonumber \\
T^{\rho \mu } &=&(k\cdot q_{1})g^{\rho \mu }-k^{\rho }q_{1}^{\mu }=(k\cdot
q_{2})g^{\rho \mu }-k^{\rho }q_{2}^{\mu },  \label{A3}
\end{eqnarray}
where $m_{f}$ denotes the mass of the fermion that emits the photon and $%
m_{f^{\prime }},$ its isospin partner in the loop. Note that the last
equality in (\ref{A3}) supposes the transversality condition $\varepsilon
\cdot k=0.$

The following identities are easily verified:

\begin{eqnarray}
ck_{\mu }\sum_{f, \mbox{colours}}Q_{f}V_{f}^{\alpha \beta \mu }(q_{1}) &=&%
\mbox{Im}\, \Pi ^{\alpha \beta }(q_{1}),  \nonumber \\
ck_{\mu }\sum_{f, \mbox{colours}}Q_{f}\widetilde{V}_{f}^{\alpha \beta \mu
}(q_{2}) &=&-\mbox{Im}\, \Pi ^{\alpha \beta }(q_{2}),
\end{eqnarray}
with $c$ defined in Eq.(\ref{3.2}).

\medskip

\noindent {\rm (ii)} {\em Axial-vector contributions}

\medskip

To obtain the general expressions for the axial-vector coefficients in Eq.(%
\ref{4.3}), we consider again the cut diagrams characterized by the momenta $%
q_{1}$ (Figs. 1a, 1c and 1e) and $q_{2}$ (Figs. 1b, 1d and 1f). Finally we
find ($f=L,U,D)$: 
\begin{eqnarray}
A_{f}^{\alpha \beta \mu }(q_{1}) &=&2\left( -m_{f}^{2}\ln \left( \frac{%
1+\upsilon _{1}}{1-\upsilon _{1}}\right) +2\upsilon _{1}E_{1}^{2}\right)
\left\{ \varepsilon ^{\alpha \beta \mu \sigma }q_{1\sigma }+\left(
\varepsilon ^{\alpha \beta \rho \sigma }q_{1}^{\mu }-\varepsilon ^{\beta \mu
\rho \sigma }k^{\alpha }\right) \frac{q_{1\rho }k_{\sigma }}{k\cdot q_{1}}%
\right\}  \nonumber \\
&&+\frac{4E_{1}\upsilon _{1}}{q_{1}^{2}}\left( E_{1}-\sqrt{q_{1}^{2}}\right)
\left\{ q_{1}^{2}\varepsilon ^{\alpha \beta \mu \sigma }k_{\sigma
}-2\varepsilon ^{\beta \mu \rho \sigma }q_{1\rho }k_{\sigma }q_{1}^{\alpha
}\right\}  \label{A4}
\end{eqnarray}

\begin{equation}
\widetilde{A}_{f}^{\alpha \beta \mu }(q_{2})=-A_{f}^{\beta \alpha \mu
}(-q_{2}).  \label{A5}
\end{equation}

If $q_{1}^{2},q_{2}^{2}>(m_{f}+m_{f^{\prime }})^{2}$ ( {\it i.e.} if both
cuts contribute to the absorptive part) then we obtain a simplified
expression for the total axial-vector contribution:

\begin{eqnarray}
A_{f}^{\alpha \beta \mu }(q_{1})+\widetilde{A}_{f}^{\alpha \beta \mu
}(q_{2}) &=&\left( q_{1}^{2}\varepsilon ^{\alpha \beta \mu \sigma }k_{\sigma
}-2\varepsilon ^{\beta \mu \rho \sigma }q_{1\rho }k_{\sigma }q_{1}^{\alpha
}\right) \frac{a_{12}}{k\cdot q_{1}}  \nonumber \\
&&-\left( q_{2}^{2}\varepsilon ^{\alpha \beta \mu \sigma }k_{\sigma
}+2\varepsilon ^{\alpha \mu \rho \sigma }q_{2\rho }k_{\sigma }q_{2}^{\beta
}\right) \frac{a_{21}}{k\cdot q_{2}},  \label{A6}
\end{eqnarray}
where
\begin{eqnarray}
a_{12} &=&m_{f}^{2}\ln \left( \frac{(1+\upsilon _{2})(1-\upsilon _{1})}{%
(1-\upsilon _{2})(1+\upsilon _{1})}\right) +2\left( \upsilon
_{1}E_{1}^{2}-\upsilon _{2}E_{2}^{2}\right)   \nonumber \\
&&+\frac{4E_{1}\upsilon _{1}(k\cdot q_{1})}{q_{1}^{2}}\left( E_{1}-\sqrt{%
q_{1}^{2}}\right) ,  \label{A7} \\
a_{21} &=&a_{12}\;(q_{1}\leftrightarrow -q_{2});  \nonumber
\end{eqnarray}
$E_{1,2}$ and $\upsilon _{1,2}$ are defined in Eqs.(\ref{A3}). In deriving
Eq.(\ref{A6}) we have used the following cyclic identity for an arbitrary
four-vector $a^{\mu }$:
\begin{equation}
\varepsilon ^{\alpha \beta \rho \sigma }a^{\mu }+\varepsilon ^{\beta \rho
\sigma \mu }a^{\alpha }+\varepsilon ^{\rho \sigma \mu \alpha }a^{\beta
}+\varepsilon ^{\sigma \mu \alpha \beta }a^{\rho }+\varepsilon ^{\mu \alpha
\beta \rho }a^{\sigma }=0.
\end{equation}
\medskip 

Notice also that in the limit of massless fermions (and with $q_{1,2}^{2}>0)$
$\upsilon _{1,2}=1,E_{1,2}=$ $\frac{1}{2}\sqrt{q_{1,2}^{2}}$ and we obtain $%
a_{12}=a_{21}=0.$ Thus, the axial-vector contribution vanishes for any
doublet of massless fermions.

\newpage

\begin{figure}[t]
\leavevmode
\par
\begin{center}
\mbox{\epsfxsize=15.cm\epsfysize=15.cm\epsffile{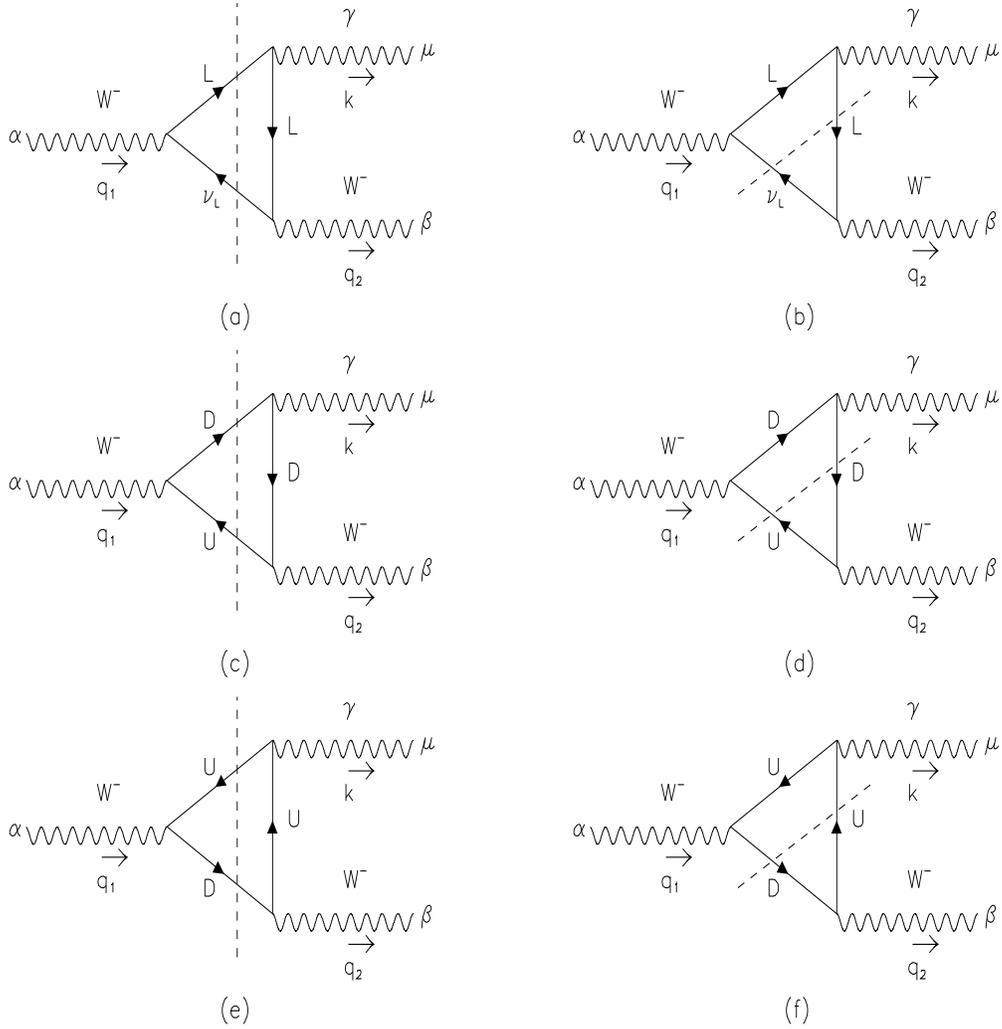}}
\end{center}
\caption{ Cut diagrams for the one-loop fermionic corrections to the $%
WW\gamma $ vertex; $L,U,D$ denote the charged leptons, up and down quarks,
respectively. }
\end{figure}

\end{document}